\documentclass[reprint,aps,graphicx,twocolumn]{revtex4}

\usepackage{amsmath} 
\usepackage{mathrsfs}
\usepackage[cmintegrals]{newtxmath}
\usepackage{graphicx}
\usepackage{dcolumn}
\usepackage{hyperref}
\usepackage{bm}
\usepackage[mathlines]{lineno}
\usepackage{color}

\begin{document}

\title{Scalable higher-order exceptional surface with passive resonators}

\author{Hong Yang$^{1}$, Xuan Mao$^{1}$, Guo-Qing Qin$^{1}$, Min Wang$^{4}$, Hao Zhang$^{1}$\footnote{zhanghaoguoqing@163.com}, Dong Ruan$^{1}$, and Gui-Lu Long$^{1,2,3,4,5}$\footnote{gllong@tsinghua.edu.cn}}

\address{$^1$State Key Laboratory of Low-Dimensional Quantum Physics, Department of Physics, Tsinghua University, Beijing 100084, China\\
$^2$Frontier Science Center for Quantum Information, Beijing 100084, China\\
$^3$Beijing National Research Center for Information Science and Technology, Beijing 100084, China\\
$^4$Beijing Academy of Quantum Information Sciences, Beijing 100193, China\\
$^5$School of Information, Tsinghua University, Beijing 100084, China
}

\date{\today}

\begin{abstract}
The sensitivity of perturbation sensing can be effectively enhanced with higher-order exceptional points due to the nonlinear response to frequency splitting. However, the experimental implementation is challenging since all the parameters need to be precisely prepared. The emergence of exceptional surface (ES) improves the robustness of the system to the external environment, while maintaining the same sensitivity. Here, we propose the first scalable protocol for realizing photonic high-order exceptional surface with passive resonators. By adding one or more additional passive resonators in the low-order ES photonic system, the 3- or arbitrary N-order ES is constructed and proved to be easily realized in experiment. We show that the sensitivity is enhanced and experimental demonstration is more resilent against the fabrication errors. The additional phase-modulation effect is also investigated.

\end{abstract}

\maketitle

\section{Introduction}


Exceptional points (EPs), arising in non-Hermitian systems, are peculiar singularities that eigenvalues and the corresponding eigenvectors simultaneously coalesce\cite{1ElGanainy2018NonHermitianPA,2zdemir2019ParitytimeSA,3Feng2017NonHermitianPB,4Miri2019ExceptionalPI,Peng2014ParitytimesymmetricWM}. The systems evolving at or around EPs exhibit richer physics\cite{6Zhang2020SyntheticAS,7zhong2020exceptional,Doppler2016DynamicallyEA,8Xu2016TopologicalET,9Wang2019ElectromagneticallyIT,ding2016emergence,zhang2018dynamically,hassan2017dynamically}, such as asymmetric mode switching\cite{Doppler2016DynamicallyEA}, topological energy transfer\cite{8Xu2016TopologicalET} and sensing\cite{12Chen2017ExceptionalPE,21Hodaei2017EnhancedSA}. Different from the linear response to external perturbation of diabolic point\cite{Zhu2010OnchipSN,zhang2017far,Li2014SingleND,Xie2021PhasecontrolledDR}, the eigenvalue splitting around EP is nonlinear following an $n$th-root dependence. Benefiting from the enhanced sensitivity to smaller perturbations, second-order EPs (EP2s) have been exploited in a wide range of sensing applications \cite{10Wiersig2014EnhancingTS,11Wiersig2016SensorsOA,12Chen2017ExceptionalPE,13Lai2019ObservationOT}. Whereas the majority of studies have focused on isolated EPs in the parameter space, which is challenging for fabrication since low tolerances are required. To overcome this difficulty, exceptional surface (ES), a two-dimensional surface of EPs, has been introduced theoretically \cite{14Zhong2019SensingWE,16Zhou2019ExceptionalSI,17Okugawa2018TopologicalES,18Budich2018SymmetryprotectedNP} and demonstrated in experiment \cite{19Zhang2019ExperimentalOO,20Qin2020ExperimentalRO}. The emergence of ES allows that experimental uncertainties do not move the system away from the EP as the sensor operates along the parameter surface. The system is robust, at the same time, enhanced sensitivity can be achieved with flexible experimental parameters.

Besides only EP2s, relative theoretical schemes\cite{22Zhong2020HierarchicalCO,23Jing2017HighorderEP,24Laha2020ThirdorderEP,Habler2020HigherorderEP,Zhang2020HighorderEP,Znojil2018ComplexSH,Xiao2019AnisotropicEP,Pan2019HighorderEP,Zhong2018PowerlawSO,Kullig2019HighorderEP} and experimental realizations\cite{21Hodaei2017EnhancedSA,Wang2019ArbitraryOE,Yu2020HigherorderEP,Zeng2019EnhancedS,Chao2020HighOrderPS,Xiao2019EnhancedSA,zhou2018optical,Zhang2019HigherorderEP} have been raised to investigate higher-order EPs, owing to the $n$th-root response near an $n$th-order EPs is extremely sensitive to the external perturbation, in contrast to the square-root response near an EP2. Higher sensitivity requires more demanding experimental conditions, all parameters have to be precisely controlled to prepare the system on EPs. In the third-order EPs implementation \cite{21Hodaei2017EnhancedSA}, three microrings are included in the system, the side cavities experience an exactly balanced gain and loss, whereas the middle one is
neutral, the resonant frequencies of the three rings have to be fine-tuned to the same value, and the gain/loss contrast has to match the coupling coefficient between the three resonators. Thus a practical and robust higher-order ES scheme against the fabrication errors and experimental uncertainties is urgently needed. 

 In this paper, we present a structure that combines the high sensitivity of the higher-order EPs with the robustness to experimental uncertainties of ES, which is originating from the structural asymmetry. Starting with the ES of order 3, our scheme is composed of two passive whispering gallery mode (WGM) resonators\cite{Cai2000ObservationOC,Wan2018ExperimentalDO,Dong2012OptomechanicalDM} and a feedback fiber waveguide, forming an unidirectional coupling between two counterpropagating modes. By monitoring the splitting of the eigenfrequencies of the systems, we find that frequency response exhibits cube-root behavior on induced small perturbations and is robust to the coupling strength between resonator and fiber. On this basis, arbitrary N size ES can be constructed by adding N-1 resonator. This proposal, in addition to naturally on the ES that can improve the stability of the system, has a number of remarkable advantages: (1) no gain resonators is used, greatly reducing the difficulty of sample fabrication; (2) the indirectly coupled resonators through optical fibers is much more realizable than direct coupling experimentally; (3) the scalability of this structure is outstanding and can contain any number of resonators with different structures.


\begin{figure}[!ht]
    \begin{center}
    \includegraphics[width=7.5cm,angle=0]{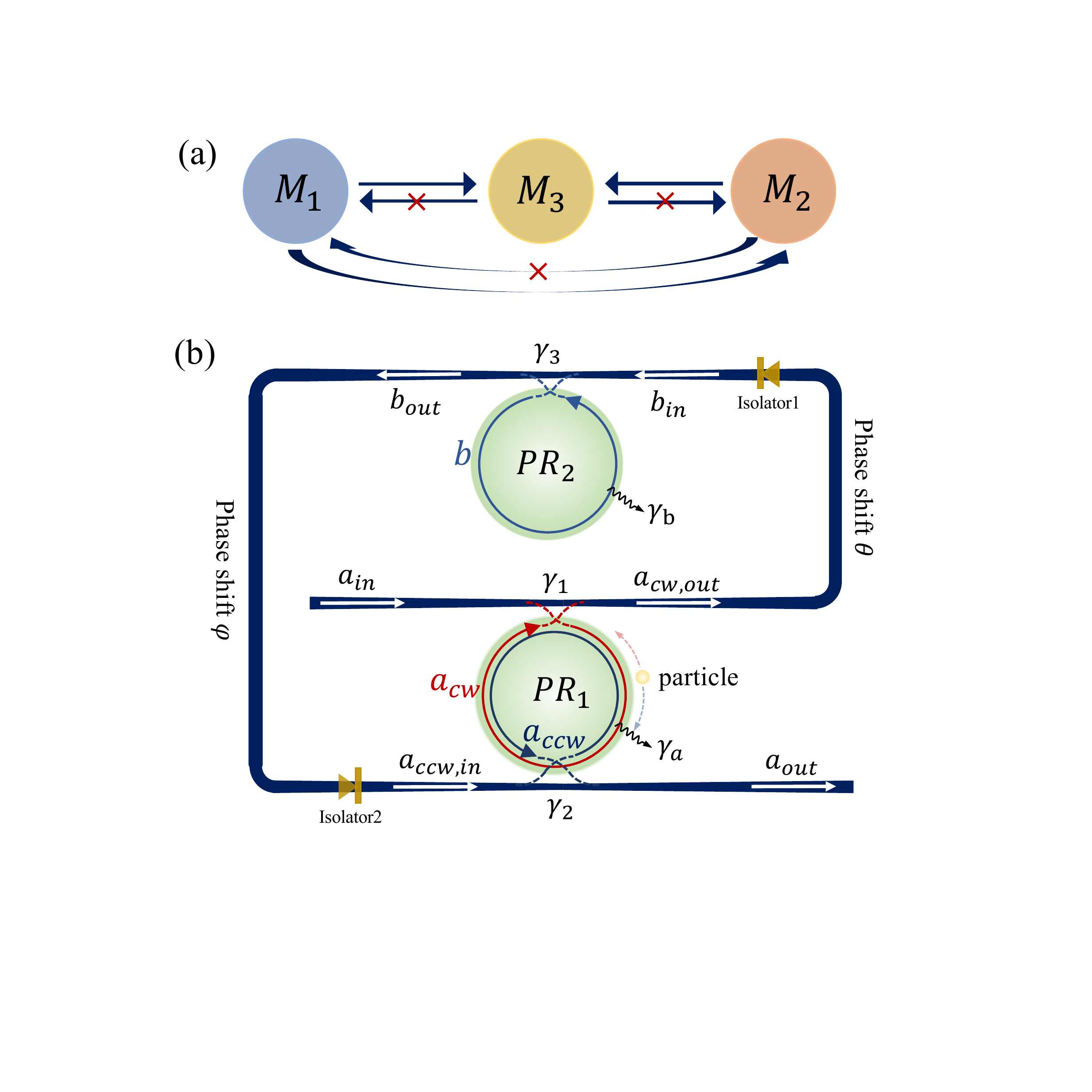}
    \caption{Schematic diagram of the third-order ES. (a) Schematic of unidirectional interaction between three modes. (b) Proposed ES coupling photonics structure consisting of two Whispering Gallery Mode (WGM) resonators and three tapered fiber couplers. The clockwise mode $a_{cw}$ and counterclockwise mode $a_{ccw}$ in lower passive resonator $PR_1$ are two degenerate resonant modes, coupling to the fiber waveguide twice by evanescent field. The WGM $b$ in upper passive resonator $PR_2$ couples with the same fiber. The light flows with different traveling directions are marked by arrow lines. To suppress the backward direction light, the optical isolator1 and isolator2 could be included. 
    } \label{fig1}
    \end{center} 
\end{figure}

\section{Model} \label{sec2}

we start with a model that can demonstrate a third-order exceptional surface to illustrate the basic principle and uncover the critical factor in achieving arbitrary order EPs. To this end, we are firstly supposed to construct a Hamiltonian that exhibits naturally degenerate eigenvalues and eigenvectors without additional request. Considering the simplest and most direct matrix, triangular matrix, having all-zero elements above or below the main diagonal, thus the eigenvalue of the matrix is the diagonal element, and setting them to the same complex value guarantees the degeneracy of the eigenvalues and eigenvectors of the system. Furthermore, non-zero elements are asked to closely relate with the parameters in system to span a two or multi- dimensional spaces. For a three-level photonic systems, a way to achieve such Hamiltonians is that the modes couple unidirectionally as shown in Fig.~\ref{fig1}(a). Mode $M_{2}$ can be excited unidirectionally by mode $M_{1}$, meanwhile, mode $M_{3}$ is motivated by mode $M_{1}$ and mode $M_{2}$, and vice versa.

Applying the above mathematical constructs to practical photonic structures, and based on an experimental realization of 2-order ES\cite{20Qin2020ExperimentalRO}, we present a third-order ES structure that easier to build. The coupling system is shown in Fig.~\ref{fig1}(b), where lower and upper WGM passive resonators $PR_1$ and $PR_2$ are coupled to a waveguide, respectively. Resonator $PR_1$, supporting a pair of degenerate clockwise (CW) and counterclockwise (CCW) resonant modes $a_{cw}$ and $a_{ccw}$ with the same eigenfrequencies $\omega_c$ and intrinsic loss rate $\gamma_a$, couples with the taper fiber twice with coupling strengths $\gamma_{1}$ and $\gamma_{2}$. For resonator $PR_2$, it is a WGMs resonator with the intracavity mode fields $b$ and resonant frequency $\omega_c$, coupling to the same fiber waveguide by the coupling strengths $\gamma_{3}$. In the absence of any disturbance, CW and CCW mode of the resonator $PR_1$ have no direct interaction. The input field $a_{in}$ is injected from the upper point of the resonator $PR_1$ to excite the CW mode, then the output field $a_{cw,out}$ acts as the input light of mode $b$ in $PR_2$ after accumulating phase shift when light propagates in the fiber. Here, an optical isolator is inserted to avoid the backscattered light. Then the joint output of mode $a_{cw}$ and $b$, after traveling along the waveguide, serves as the input field to excite the CCW mode of resonators $PR_1$. The isolator 2 prohibits another path to excite the CCW mode $a_{ccw}$, i,e., the CW mode $a_{cw}$ couples into the waveguide in the lower joint part, propagating in the waveguide, couples into the resonator again in the upper joint part to excite the CCW mode $a_{ccw}$. The absence of this excitation path forms the asymmetric coupling between the $a_{cw}$ and $a_{ccw}$ optical modes. Since the third-order ES system constructed is used for perturbation detection, there are two degenerate and counterpropagation modes in the same resonator. The isolator 1 achieves the unidirectional excitation of mode $a_{cw}$ to mode $b$, the introduction of isolator 2 and asymmetric waveguide structure is the key of another unidirectional excitation of mode $a_{cw}$ and $b$ to mode $a_{ccw}$, which is consistent with our original idea.

The dynamics of the system can be described by the coupled mode theory given by
\begin{align}
   \frac{\mathrm{d}a_{cw}}{\mathrm{d}t}=&(-i\omega_c-\Gamma_a) a_{cw}-\sqrt{\gamma_1} a_{in},\\
   \frac{\mathrm{d}b}{\mathrm{d}t}=&(-i\omega_c-\Gamma_b) b-\sqrt{\gamma_1\gamma_3}e^{i\theta} a_{cw}-\sqrt{\gamma_3}e^{i\theta} a_{in},\\
   \frac{\mathrm{d}a_{ccw}}{\mathrm{d}t}=&(-i\omega_c-\Gamma_a)a_{ccw}-\sqrt{\gamma_1\gamma_2}e^{i(\theta+\varphi)} a_{cw}\nonumber\\&- \sqrt{\gamma_2\gamma_3}e^{i\varphi} b-\sqrt{\gamma_2}e^{i(\theta+\varphi)} a_{in},
\end{align}
where $\Gamma_a=\frac{\gamma_{a}+\gamma_1+\gamma_2}{2}$, $\Gamma_b=\frac{\gamma_{b}+\gamma_3}{2}$, and $\gamma_b$ is the intrinsic loss of $PR_2$, angle $\theta$ and $\varphi$ is the phase shift propagating along the waveguide between the two resonators. The phase induced by the waveguide arm can be written as $\theta(\varphi)=2\pi\frac{nL_{\theta(\varphi)}}{\lambda}$, $n$ is the effective refractive index of the waveguide and $L_{\theta(\varphi)}$ is the length of the fiber between the two resonators.

In the absence of the input field, we can obtain the effective coupled-mode equations. In this case, the system can be described by 
$i\hbar\frac{\mathrm{d} \vec{V}}{\mathrm{d}t}=\mathbf{H_0}\vec{V},$
where $\vec{V}=(a_{cw},b,a_{ccw})^{T}$ represents the field amplitudes of two resonators, $\mathbf{H}_0$ is the associated $3\times3$ non-Hermitian Hamiltonian and can be written as
\begin{align}
\mathbf{H_0}=\begin{pmatrix} 
\omega_c-i\gamma & 0 & 0 \\ \mu_{1} &  \omega_c-i\gamma & 0 \\ \mu_{2} & \mu_{3} & \omega_c-i\gamma
\end{pmatrix},
\end{align}
where $\mu_{1}=-i\sqrt{\gamma_{1}\gamma_{3}}e^{i\theta}, \mu_{2}=-i\sqrt{\gamma_{1}\gamma_{2}}e^{i(\theta+\varphi)}$ and $\mu_{3}=-i\sqrt{\gamma_{2}\gamma_{3}}e^{i\varphi}$. These three non-zero parameters of diagonal matrix are the function of $\gamma_1$, $\gamma_2$, and $L$ under the constraint condition $\gamma=\Gamma_a=\Gamma_b$ with certain value of $\gamma_a$ and $\gamma_b$. Without perturbation, the Hamiltonian $\mathbf{H_0}$ of our system spontaneously exhibits the 3$\times$3 triangular normal form, all three eigenfrequencies coalesce at $\omega_{n}$ = 0, n={0,1,2}.

\begin{figure}[!ht]
    \begin{center}
    \includegraphics[width=8.5cm,angle=0]{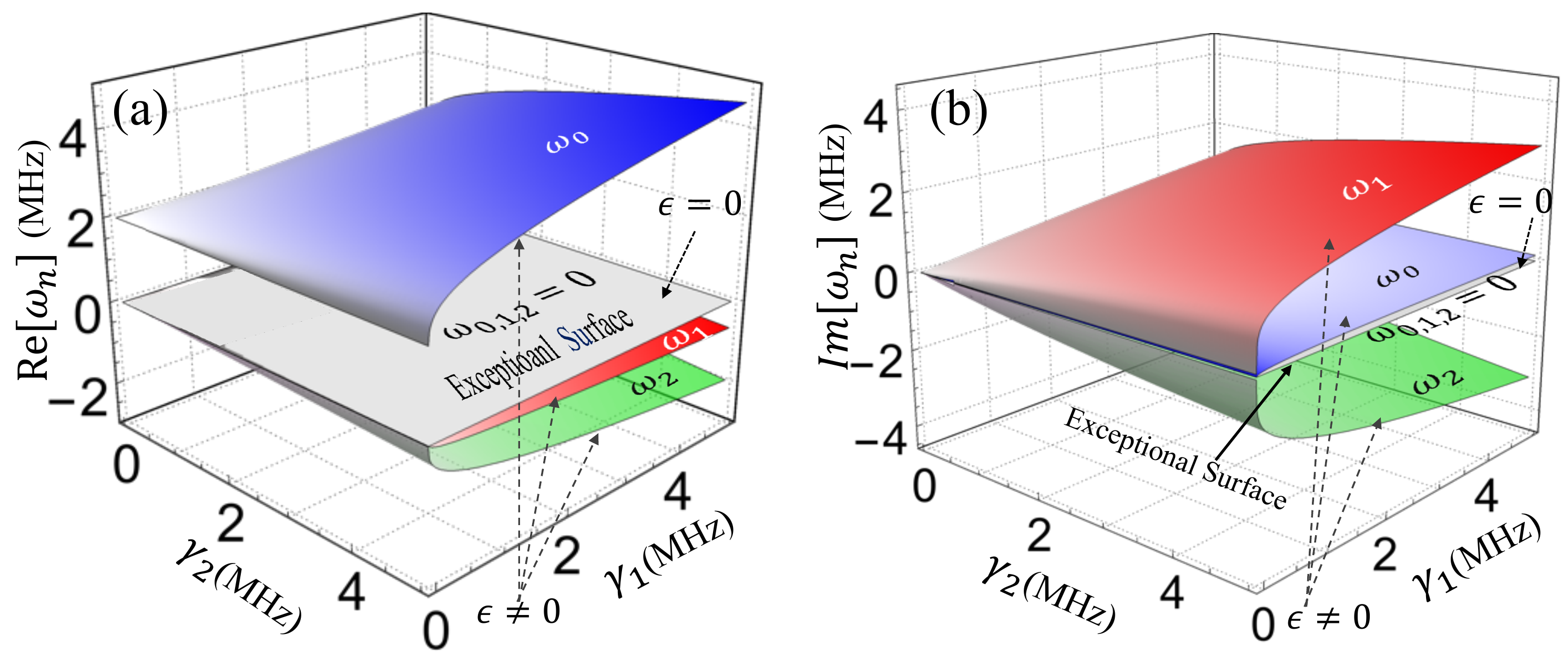}
    \caption{Complex eigenfrequencies around a third order exceptional surface in the parameter sapce of [$\gamma_1, \gamma_2$]. (a) The real parts of the eigenfrequencies. (b) The imaginary parts of the eigenfrequencies. The gray plane shows that when there is no perturbation, an ES is formed in the parameter space of [$\gamma_1, \gamma_2$], the eigenvalue is coalesced. The appearance of perturbation destroys the degeneracy of eigenvalues and the frequency splitting occurs. The parameters The parameters are $\gamma_a=\gamma_b$, $\gamma_3=\gamma_1+\gamma_2$, $\theta=\frac{2\pi}{5}$, $\varphi=\frac{\pi}{2}$ and $\epsilon=$1 MHz.} \label{fig2}
    \end{center} 
\end{figure}

When a perturbation is applied to resonator $PR_1$, leading to the frequency shift and coupling of mode $a_{cw}$ and $a_{ccw}$, it is given by an off-diagonal Hamiltonian\cite{22Zhong2020HierarchicalCO},
\begin{align}
\mathbf{H}_{1}=\begin{pmatrix} 
\epsilon & 0 & \epsilon \\ 0 & 0 & 0 \\ \epsilon & 0 & \epsilon
\end{pmatrix}.
\end{align}
The perturbed Hamiltonian is $\mathbf{H}=\mathbf{H_0}+\mathbf{H_1}$, the characteristic equation on the basic of $\omega_c-i\gamma$ is
\begin{align}
\omega_n^{3}-2\epsilon\omega_n^2-\mu_{2}\epsilon\omega_n-\mu_{1}\mu_{3}\epsilon=0,
\end{align}
the roots of this cubic equation can be solved analytically and 
our analysis indicates that the real parts of $\omega_0$
and $\omega_2$ diverge from each other in an $\epsilon^{\frac{1}{3}}$ fashion. Consequently, the eigenfrequencies of the system can be assessed by monitoring the transmission spectral lines, obtained from the cavity input-output relation,
\begin{align}
a_{out}=e^{i(\theta+\varphi)}(a_{in}+\sqrt{\gamma_1}a_{cw}) + e^{i\varphi}\sqrt{\gamma_3} b + \sqrt{\gamma_2} a_{ccw},
\end{align}
the corresponding transmission spectrum is $T=|\frac{a_{out}}{a_{in}}|^2$.

\section{Result of third-order ES}
\label{sec3}

When there is no disturbance in resonator $PR_1$ ($\epsilon$=0), Hamiltonian $\mathbf{H_0}$ is initial triangular matrix, which drives the system forward to an exceptional surface in the parameter space $[\gamma_1,\gamma_2]$ as shown in gray plane in Fig.~\ref{fig2}, although the three eigenstates coalesce, the exceptional point can be distinguished one from the other through the coupling loss $\gamma_1, \gamma_2$. By introducing a small perturbation in resonator $PR_1$($\epsilon \neq$ 0), the disturbance drives the system out of the state of degeneracy and three complex eigenfrequencies are obtained. The system behaves differently with different coupling sets, that is when the system operates at different locations of the ES, the eigenfrequencies are changed depending on the different exceptional points. It is obvious that mode splitting increases along with the strength of coupling loss and the enhanced sensitivity can always be available in a large range of parameters. The ES ensures that one can operate the system around EPs with variable parameters.
  \begin{figure}[!ht]
    \begin{center}
    \includegraphics[width=8.5cm,angle=0]{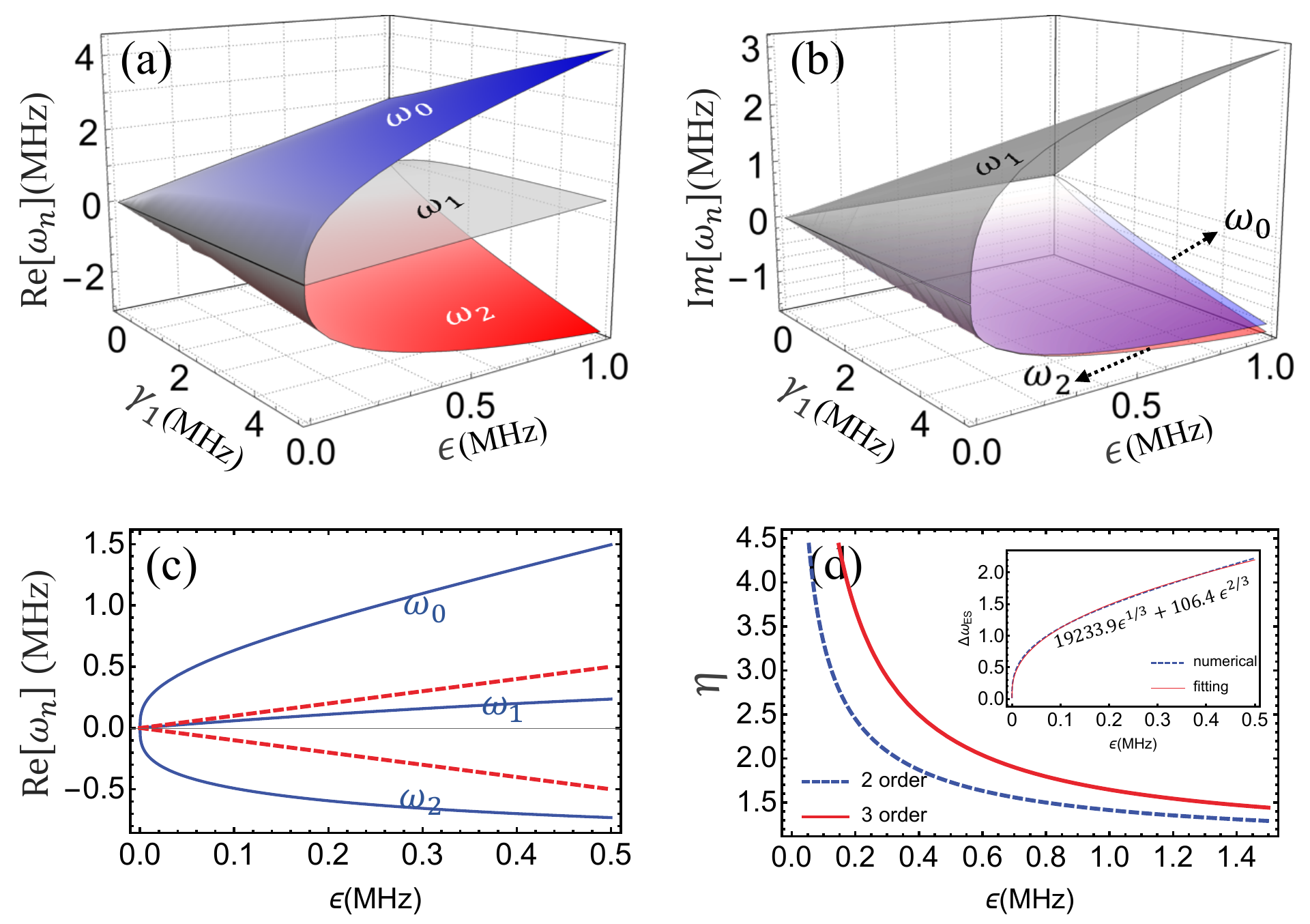}
    \caption{(a), (b) Complex eigenfrequencies splitting vs perturbation strength $\epsilon$ and coupling strength $\gamma_1$. (c) The eigenfrequencies vs the coupling rate $\epsilon$ at the DP (red dashed line) and EP (blue solid line). (d) The enhancement of the mode splitting between 2-order and 3-order ES vs $\epsilon$. The insert is numerical (dashed blue line) and fitting (solid red line) results of eigenfrequency difference, responding cube-root behavior as a function of $\epsilon$. The parameters are $\gamma_
   a=\gamma_b$, $\gamma_1=\gamma_2$, $\gamma_3=\gamma_1+\gamma_2$, $\theta=\frac{\pi}{3}$, $\varphi=\frac{\pi}{6}$, and $\gamma_1=1$ MHz in (c), (d).}\label{fig3} 
    \end{center} 
\end{figure}
\begin{figure*}[!ht]
    \begin{center}
    \includegraphics[width=15cm,angle=0]{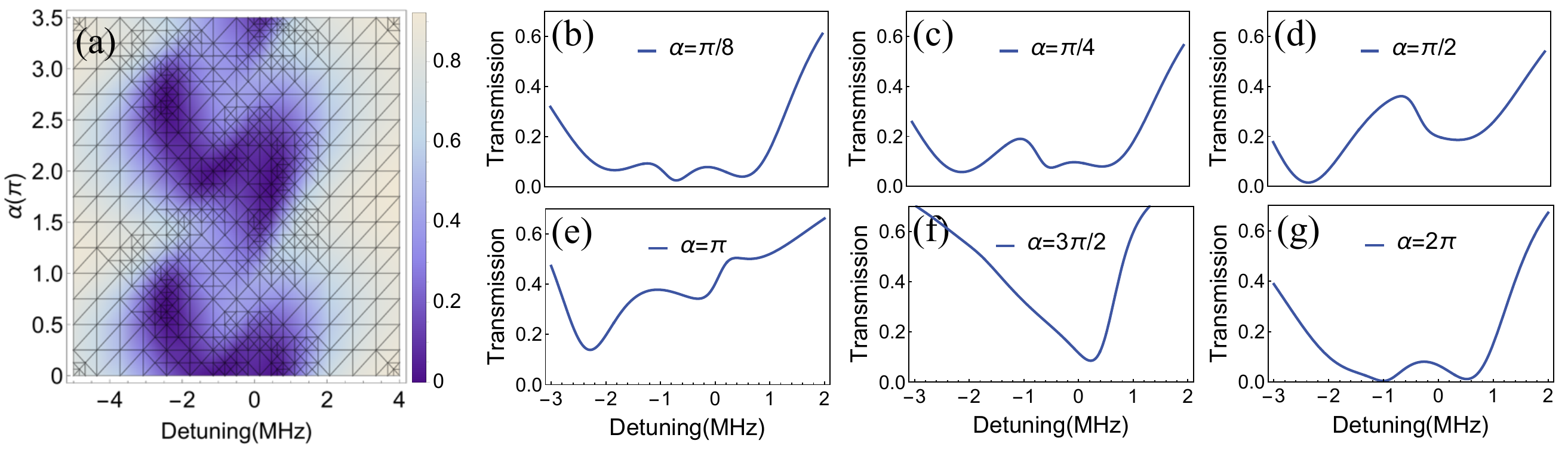}
    \caption{(a) The transmission spectrum vs the detuning and phase $\alpha$. (b)-(g) Transmission spectrum with the phase $\alpha=[\frac{1}{8}, \frac{1}{4}, \frac{1}{2}, 1, \frac{3}{2}, 2]\pi$. The parameters are $\gamma_a=\gamma_b=0.2$ MHz, $\gamma_1=\gamma_2$=1MHz, $\gamma_3=2(\gamma_1+\gamma_2)$ and $\epsilon=$1 MHz. }\label{fig4} 
    \end{center} 
\end{figure*}  
 
 To show the system reacts to a small variation in the resonator $PR_1$ around the third-order EPs, three complex eigenfrequencies as a function of $\gamma_1$ and $\epsilon$ are obtained by numerical calculation as shown in Fig.~\ref{fig3}(a), (b). As for a normal diabolic point, i.e., there is no coupling between fiber and resonator $PR_2$ and lower point of $PR_1$, the eigenfrequency splitting induced by perturbation is 2$\epsilon$, as shown in the red dashed line of Fig.~\ref{fig3}(c). when the coupling is formed, as a consequence of this perturbation, the degenerate energy levels split into three distinct branches as shown in the blue solid line of Fig.~\ref{fig3}(c), and the eigenfrequency difference between $\omega_0$ and $\omega_2$ is also plotted as a function of $\epsilon$ shown in the insert of Fig.~\ref{fig3}(d). By fitting the numerical results, we find that perturbations around a third-order exceptional point experience an enhancement of the form $\epsilon^{\frac{1}{3}}$. Here, we define a sensitivity enhancement factor $\eta=\frac{\Delta\omega_{EP}}{\Delta\omega_{DP}}$, where $\Delta\omega_{EP}$ and $\Delta\omega_{DP}$ are the eigenfrequency difference induced by perturbation of EP and DP system, respectively.   Fig.~\ref{fig3}(d) illustrates the enhancement factor $\eta$ as a function of $\epsilon$. it is notable that the enhancement factor is higher for sufficiently small perturbation. For comparison, second-order EP (blue dashed line) and third-order EP (red solid line) are plotted respectively. The external perturbation response near an EP3 is extremely more sensitive to the EP2.

Mode splitting can be obtained from the power spectra as shown in Fig.~\ref{fig4}, the theoretically expected indicates that the frequencies splitting is supposed to occur for the smaller perturbation, however, we note that this is limited by the resolvability due to the finite linewidths associated with the imaginary part of the corresponding eigenfrequency. For practical applications, the gain-assisted resonators can be used to suppress the linewidths and further improve the sensitivity\cite{Hodaei2014ParitytimesymmetricML,Liu2018GainLC}. It is worth mentioning that in addition to perturbation that can provide coupling between CW and CCW, the unique structure of our system provides an asymmetric coupling between these two counter-propagating modes, that is, CCW mode can be excited by CW mode through the feedback fiber waveguide, which provides another degree of freedom to control the frequency splitting by adjusting the phase shift $\alpha=\theta+\varphi$ as shown in Fig.\ref{fig4}. With the increase of phase shift, the shape of splitting spectra transforms from symmetric to asymmetric. Here, the factor of affecting the eigenfrequency is the total accumulated phase rather than the local one of segment. Therefore, there is no special requirement for the resonator position while ensuring that the total phase is at the desired value.

\begin{figure}[!ht]
    \begin{center}
    \includegraphics[width=8.5cm,angle=0]{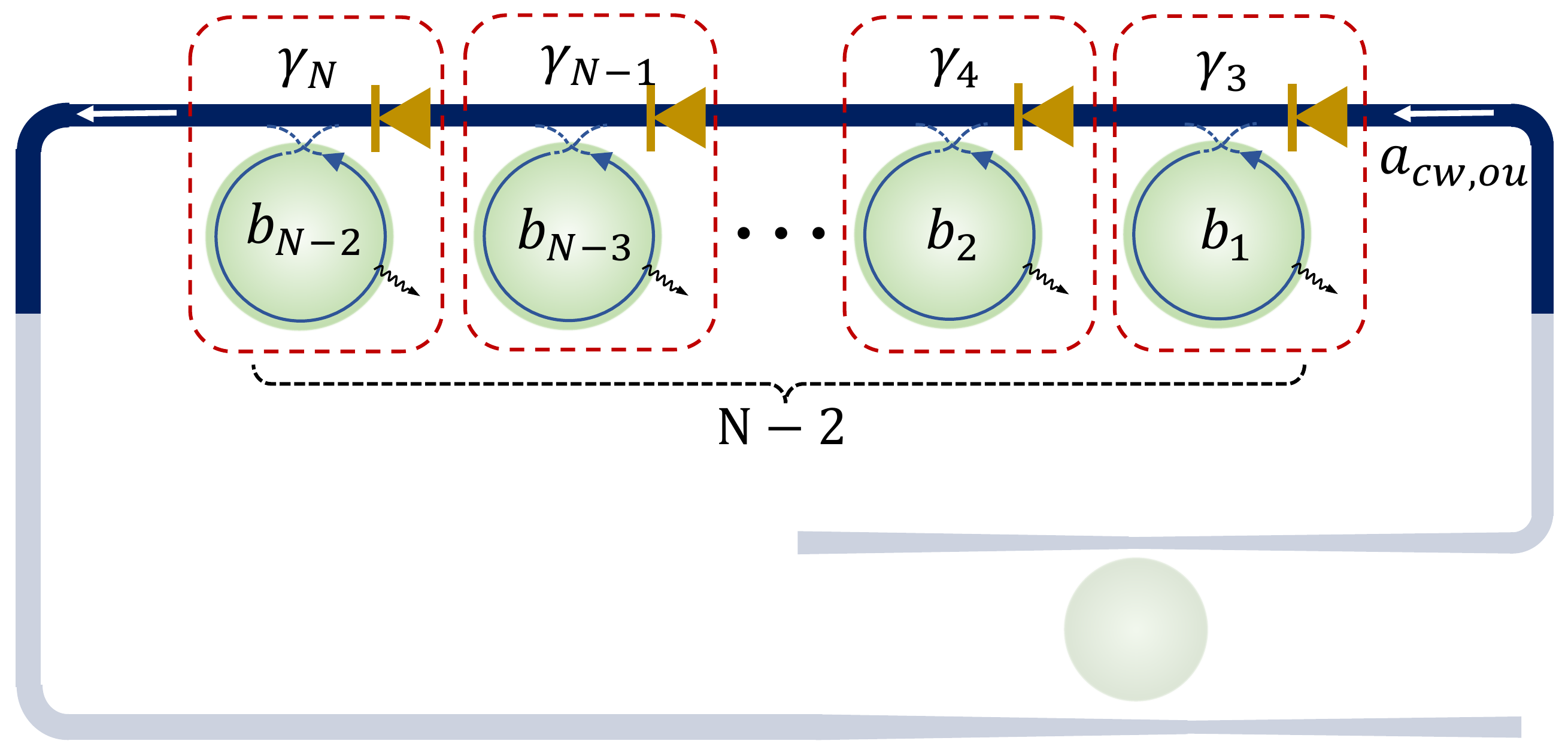}
    \caption{An N-order ES schematic introducing N-2 resonators.}\label{fig5} 
    \end{center} 
\end{figure}

\section{ES construction of arbitrary size}
\label{sec4}

To extend the third-order scheme, we gradually increase the size of the target Hamiltonian by adding the unit composed of a resonator and an isolator one by one. The above third-order ES and their behaviors are universal, higher-order ES with arbitrary size N is constructed successfully. For the N-order system shown in Fig.\ref{fig5} constructed by adding N-2 units, the corresponding Hamiltonian is given by
\begin{align}
\begin{pmatrix}
        \omega_c-i\gamma & 0 & \cdots & 0\\
        \mu_{21} & \omega_c-i\gamma & \ddots & \vdots\\
        \vdots & \ddots & \ddots & 0\\
        \mu_{n1} & \cdots & \mu_{n,n-1} & \omega_c-i\gamma\\
    \end{pmatrix},
\end{align}
the N$\times$N triangular matrix has N degeneracy eigenvalues, forming an N-order ES in the parameter space. For the perturbation $\epsilon$ acting on the resonator, the eigenfrequency response is proportional to $\epsilon^{\frac{1}{N}}$, indicating that the sensitivity enhancement is higher for the smaller perturbation. 

Each small unit is indirectly coupled together through fiber waveguide, avoiding the influence of the newly added unit on the previous one. Comparing to the direct coupling ways, our proposal is more flexible and stable in experiment. Furthermore, there is less limit to the type of resonators used in each unit, it can be a microdisk cavity, a microsphere cavity, a microring cavity, and etc., which are selected according to the requirement of the experiment, and only need to ensure that the resonance frequency is consistent. Moreover, the position of the resonators can be selected at will, as long as the phase accumulated by the total fiber length is the target phase.


\section{Conclusion}\label{sec5}

In summary, we introduce a new approach to construct a higher-order ES scheme for improving the robustness of the system against external disturbances. Starting from the Hamiltonian construction, we obtained the necessary conditions for obtaining the higher-order ES. On this basis, by designing the photonic structure, the third-order ES has been demonstrated in the two-dimensional parameter space composed of the coupling coefficients between the detecting resonator and the fiber. Theoretical analysis shows that the energy level splitting of the system is proportional to the cube root of the perturbation. Compared with the second-order ES and traditional DP, the sensitivity is greatly improved. We also explore the transmission spectrum and investigate the modulation of the spectrum shape by varying the additional phase caused by the waveguide structure. Adding resonators one by one, the third-order scheme can be easily extended to arbitrary N order, depending only on the number of resonators and the positions of the resonators do not affect the order. Since no gain resonators are used and indirectly coupled applied to all resonators by fiber waveguide, our scheme is easier to build experimentally.

\section*{ACKNOWLEDGMENT}
The work was supported by the National Natural Science Foundation of China under Grants (61727801); National Key Research and Development Program of China (2017YFA0303700); The Key Research and Development Program of Guangdong province (2018B030325002)..

\bibliography{ES}


\end{document}